# THE INTERDEPENDENCE OF HIERARCHICAL INSTITUTIONS: FEDERAL REGULATION, JOB CREATION, AND THE MODERATING EFFECT OF STATE ECONOMIC FREEDOM

**David S. Lucas**
Whitman School of Management
Syracuse University
721 University Ave
Syracuse, NY 13244 USA
Tel: 1-410-852-9166
Email: dlucas01@syr.edu

**Christopher Boudreaux**
College of Business
Florida Atlantic University
777 Glades Road, KH 145
Boca Raton, FL 33431 USA
Tel: 1-561-297-3221
Email: cboudreaux@fau.edu

**Abstract.** Regulation is commonly viewed as a hindrance to entrepreneurship, but heterogeneity in the effects of regulation is rarely explored. We focus on regional variation in the effects of national-level regulations by developing a theory of hierarchical institutional interdependence. Using the political science theory of market-preserving federalism, we argue that regional economic freedom attenuates the negative influence of national regulation on net job creation. Using U.S. data, we find that regulation destroys jobs on net, but regional economic freedom moderates this effect. In regions with average economic freedom, a one percent increase in regulation results in 14 fewer jobs created on net. However, a standard deviation increase in economic freedom attenuates this relationship by four fewer jobs. Interestingly, this moderation accrues strictly to older firms; regulation usually harms young firm job creation, and economic freedom does not attenuate this relationship.

# 1. Introduction

Scholars in the tradition of New Institutional Economics (Acemoglu, Johnson, & Robinson, 2005; North, 1990; Williamson, 2000) have shown increasing interest in the influence of formal rules and informal social rules on entrepreneurship (Bruton, Ahlstrom, & Li, 2010; Bylund & McCaffrey, 2017; Klein, Mahoney, McGahan, & Pitelis, 2017). The consistent insight emerging from this view is that entrepreneurial activity tends to yield productive outcomes when the institutional framework is favorable (Bjørnskov & Foss, 2016; Bradley & Klein, 2016). When institutions incentivize productive courses of entrepreneurial action at the micro level (Baumol, 1990; Murphy, Shleifer, & Vishny, 1991; Sobel, 2008), employment and job creation result at the macro-level (Baumol & Strom, 2007; Bjørnskov & Foss, 2013; Murphy, Shleifer, & Vishny, 1993). On the other hand, prohibitive institutions tend to reduce the economic benefits of entrepreneurship. Regulation, for instance, is a key barrier to venture creation and growth. Researchers have argued that regulation reduces industry employment growth (Bailey & Thomas, 2017; Bertrand & Kramarz, 2002), creates barriers to new firm entry (Djankov, La Porta, Lopez-de-Silanes, & Shleifer, 2002), hinders investment (Escribá-Pérez & Murgui-García, 2017), and asymmetrically burdens small firms (Bradford, 2004; Klapper, Laeven, & Rajan, 2006).

Despite its consistency, the entrepreneurship literature on regulation tends to present a simplistic view in that it fails to account for heterogeneous effects of regulatory institutions (Kim, Wennberg, & Croidieu, 2016). Many new ventures *do* emerge and grow in highly regulated industries, and the young, high-growth firms that drive job creation exist in all types of industries under widely varying regulatory regimes (Henrekson & Johansson, 2010). Furthermore, regional variation in entrepreneurial activity within a national institutional framework suggests the possibility of further heterogeneity of the effects of those national institutions (Fritsch & Storey, 2014). Why might national institutions like regulation be more constraining for entrepreneurial job creation in some regions than in others?

To answer this question, we develop a theory of *hierarchical institutional interdependence*: the effects of national institutional rules are contingent on the regional institutional framework. Specifically, we argue that regional economic freedom moderates the negative impact of national regulation for a key entrepreneurial outcome for economic growth: net job creation[1]. Recognizing not only the *existence of* but also the *relationship among* multiple, hierarchically situated governing units is a critical advance in our understanding of the institutions-entrepreneurship relationship. By explicitly modeling and measuring formal institutions at both the national and regional levels, we can more accurately depict the institutional environment, and we can assess both the direct and indirect consequences of institutions on economic growth.

We examine why the effects of national regulation on entrepreneurship vary across regions. One underexplored possibility is that regional policies—especially those that encourage the creation of private enterprises— might moderate the effects of national regulation. We hypothesize that regional economic freedom offsets the costs of national regulation on job creation. Our theory integrates insights from New Institutional Economics (Acemoglu et al., 2005; North, 1990; Williamson, 2000) and public choice theories of regulation (Peltzman, 1976; Posner, 1974; Stigler, 1971) with the political science theory of market-preserving federalism (Weingast, 1995). Market preserving federalism suggests that regional policymakers possess more "local knowledge" about their particular economic context, and so they are able to oppose and counteract federal rules that do not accord with regional enterprise needs. We build on this to theorize that regional economic freedom can be a mechanism for countervailing the effects of extant national regulation on entrepreneurial activity.

---

[1] Net job creation = job gains – job losses. We say that net jobs are created if job gains > job losses, but net jobs are destroyed if job gains < job losses. This is important because the economy grows with higher net job creation. This is explained in more detail in section 3.

Our key outcome is net job creation in an industry. This measure speaks to the key economic benefits of entrepreneurial activity, as it the emergent outcome of Schumpeterian innovation and competitive processes and is largely driven by entrepreneurial ventures (Birch, 1987; Decker, Haltiwanger, Jarmin, & Miranda, 2014). One additional advantage of this measure is that we can examine the effects of regulation on entrepreneurs relative to incumbents, by comparing results across the firm age spectrum. Although young firms are simultaneously burdened by regulation (Stigler, 1971) and enabled by a favorable regional context (Audretsch, Falck, Feldman, & Heblich, 2012), few studies directly compare the effects of institutions on new versus incumbent firm activity. We address this in our analysis, hypothesizing that local and regional policies that offset national regulation will have a greater effect on young firms rather than incumbents. Lastly, we address the call to more closely examine the heterogeneity in the effect of national-level regulation on entrepreneurship at the local level (Audretsch, Belitski, & Desai, 2018). By analyzing how the effects of regulation vary across regions and between different aged-firms, our study suggests the relationship between regulation and entrepreneurship is more nuanced than previously considered (Dilli, Elert, & Herrmann, 2018).

We test our hypotheses with a three-way fixed effects model, which allows us to control flexibly for unobserved industry, county, and year-specific effects in addition to relevant regional covariates. Our results are consistent with our core hypotheses: regulation is associated with less job creation, economic freedom is associated with more job creation, and economic freedom attenuates the negative regulation-job creation relationship. Specifically, we find a one percent increase in industry-level regulation is associated with 14 fewer net jobs created in a county on average, but this effect varies with the quality of the regional institutional environment. In states that have very little economic freedom, a one percent increase in industry-level regulation is associated with roughly 27 fewer net jobs created in a county. On the other hand, regulation has no discernable effect on net job creation in states that have very high levels of economic freedom. Interestingly, and contrary to our expectations, the moderating effect appears strongest for older firms; in fact, we find no evidence of moderation for the youngest firms. We also decompose the economic freedom into its three sub-indices: fiscal freedom (government spending), tax freedom, and labor market freedom. We find that moderating effect holds for the tax and labor sub-indices but not for the fiscal freedom sub-index.

These results are important for several reasons. First, we account for the multilevel, embedded nature of institutions to develop a more nuanced understanding of their implications for job creation. This approach follows recent multilevel work that has done much to clarify how the institutional context informs entrepreneurship—for instance, by shaping the motivations, aspirations, and decision making of entrepreneurs (Autio & Acs, 2010; Boudreaux, Nikolaev, & Klein, 2018; Estrin, Korosteleva, & Mickiewicz, 2013). Studies at the regional-level or individual-level often neglect the country-level context (Charron, Dijkstra, & Lapuente, 2014; Spigel, 2016, 2017), whereas studies at the country-level (Bjørnskov & Foss, 2008; McMullen, Bagby, & Palich, 2008; Nyström, 2008) often lose nuance through aggregation. By contrast, the multilevel approach provides additional nuance that is often overlooked by researchers (Audretsch et al., 2018). And while scholars have paid much attention to national and regional enterprise policy separately, ours is one of the first studies to directly investigate the interaction among policies across institutional levels (Schröder & Voelzkow, 2016).

Second, we contribute to the literature on institutions and entrepreneurship by focusing on the disparate impacts of regulation for firm growth across the firm age spectrum. Recent work suggests that institutions influence entrepreneurial growth aspirations (Estrin et al., 2013), but we know less about the effects of regulation on actual growth outcomes, such as employment growth. Furthermore, while it is understood that new entrants differ from incumbents in growth decisions (Gilbert, McDougall, & Audretsch, 2006), the differential effects of institutions on growth outcomes across the firm age spectrum have yet to be explored.

Third, our evidence affirms regional economic freedom, in addition to national, as an important object for entrepreneurship research (McMullen et al., 2008). We extend the insight that economic freedom is beneficial for job creation to the regional context; however, we also find an underexplored benefit of regional economic freedom: its ability to offset the cost of national policy. This suggests that regional policymakers have the potential to cultivate a promising institutional framework through at least two dimensions of economic freedom, including favorable tax regimes and flexible labor market policies. Fourth, and finally, we offer the surprising finding that the costs of regulation and the offsetting benefits of regional economic freedom appear to be primarily concentrated among older firms—at least in terms of employment growth. We offer some possible explanations for this finding.

## 2. Theory and hypotheses

With the rise of New Institutional Economics as a predominant perspective on economic development, scholars have increasingly recognized that human action does not occur in an institutional vacuum (North, 1990; Williamson, 2000). The notion that the political "rules of the game" impact economic activity is now widely accepted; institutions shape the relative rewards, risk attitudes, and expectations about future outcomes (Bylund & McCaffrey, 2017; Williamson, 2000). However, the relationship between institutions and job creation specifically has received less attention—particularly from a regional perspective. Furthermore, studies that focus on the institutional determinants of job creation tend to treat policy as a single-level phenomenon, e.g., by considering only national regulation (Bailey & Thomas, 2017; Bertrand & Kramarz, 2002) or national economic freedom (Bjørnskov & Foss, 2008) without attending to spatial heterogeneity arising from regional enterprise policy.[2] Thus, before delving into the interaction of multilevel governance institutions, we develop baseline hypotheses on the direct effects of national regulation and regional economic freedom.

### 2.1. Regulation

Regulation is expected to deter job creation for a number of reasons. The first relates to the nature of regulation as outlined in the economic theory of regulation (Peltzman, 1976; Stigler, 1971). This view posits that interest groups influence the outcome of the regulatory process by providing support to politicians or regulators. Actors involved in the regulatory process are assumed to be rational and self-interested—the same as everyone else (Buchanan, 1984). Furthermore, politicians are able to extract personal benefits via the regulatory process in the form of political support and campaign contributions (Holcombe, 2002, 2013), and regulators are able to extract personal benefits like appointments to lucrative and powerful positions in the regulated sector (Holcombe & Boudreaux, 2015). When regulation is the outcome of self-interested exchange, it is said to yield concentrated benefits to organized interests while yielding dispersed costs to others—e.g., to firms now facing additional compliance costs.

---

[2] Williamson (2000) provides a framework that categorizes institutions into a four-level hierarchy. This theoretical framework has been applied entrepreneurship to better understand institutional context (Bylund & McCaffrey, 2017; Estrin, Korosteleva, & Mickiewicz, 2013; Misangyi, Weaver, & Elms, 2008; Pacheco, York, Dean, & Sarasvathy, 2010). Informal institutions (level one) are deeply embedded in society and include customs, traditions, and religious norms; they emerge spontaneously and change over a long period of time (100 to 1000 years). Formal institutions (level two of the hierarchy) relate to the efficacy of government action including property rights protection and the quality of the regulatory environment; they can change more rapidly (ten to 100 years). Governance (level three) represents the play of the game, which can change even more rapidly (one to ten years). Lastly, the fourth level is that of resource allocation, employment choices, and entrepreneurship (Williamson, 2000). Our dependent variables deal with hierarchically-structured formal institutions *within* level two of Williamson's (2000) hierarchy.

In addition to compliance costs, an extension of the economic theory of regulation to the 'revolving door hypothesis' (Blanes i Vidal, Draca, & Fons-Rosen, 2012; Gormley Jr, 1979) suggests that complex and extensive regulation is a byproduct of this regulatory phenomenon. When regulations are complex, firms need experts to help navigate through the red tape. Who better to hire than the ones who wrote the regulation? As (Schweizer, 2013, p. 17) argues, "There is money to be made in creating complex rules and laws that nobody can understand. Those who write these laws and regulations can leave their posts and charge companies large fees to decipher the very regulations they wrote." Regulatory complexity further raises the cost of hiring and retaining employees. Regulators and ex-employees of regulatory agencies can help firms navigate this onerous process, but they can also use the political process to encourage rent seeking (Krueger, 1974; Murphy et al., 1993; Tullock, 1967) and to create entry barriers (Dean & Brown, 1995; Djankov et al., 2002).

Rent seeking is additionally costly for economic growth (Murphy et al., 1993) because it encourages unproductive entrepreneurship at the expense of productive entrepreneurship (Baumol, 1990; Murphy et al., 1991; Sobel, 2008). Rent seeking entails a reallocation of efforts by incumbents from market to political competition, which is especially destructive when it results in entry barriers. For instance, rent seeking strategies may result in occupational licensing (Kleiner, 2000; Meehan & Benson, 2015) or may force new entrants to adopt more stringent regulations via "grandfather clauses" (Dean & Brown, 1995). Consistent with this logic, regulations that raise the costs of entry have been shown to hinder job creation among new firms (Bertrand & Kramarz, 2002; Branstetter, Lima, Taylor, & Venâncio, 2014) Firms can use rent seeking to persistently distort efficient resource allocation (Caves & Porter, 1977), often through a coalition of special interests (Olson, 1965).

Politically-connected firms are able to use their political capital to profit from regulatory activity (Fisman, 2001). Firms with strong political connections often curry special favors from regulatory agencies and face lower hurdles than firms with weaker political connections (Berkman, Cole, & Fu, 2010). For instance, entrepreneurship in China is governed more by political capital and less by market fundamentals (Ge, Stanley, Eddleston, & Kellermanns, 2017). If rent seeking encourages unproductive and destructive entrepreneurship, less connected firms will have to devote more resources to dealing with the political process and satisfying regulatory requirements.

In sum, ample empirical evidence suggests that regulatory restrictions impose additional compliance costs, causing firms to reallocate scarce resources away from productive growth. Consequently, we should observe less economic activity—such as net job creation—with increasing regulation. For these reasons, we propose our first hypothesis:

**Hypothesis 1:** Regulation is associated with less net job creation (or more net job destruction).

### 2.2. Regional economic freedom

Economic freedom, defined broadly as the extent to which political institutions facilitate personal choice, voluntary exchange, and market competition, has been shown to nurture entrepreneurship (Bjørnskov & Foss, 2008, 2016; Boudreaux, 2014; Boudreaux, Nikolaev, & Klein, 2017; McMullen et al., 2008; Nikolaev et al., 2018; Nyström, 2008) and economic development (Carlsson & Lundström, 2002; de Haan & Sturm, 2000; Gwartney, Lawson, &

Holcombe, 1999). High economic freedom is said to be integral to a pro-market institutional environment, encouraging productive entrepreneurship; in contrast, institutional environments characterized by low economic freedom tend to encourage unproductive and destructive entrepreneurship (Baumol, 1990; Boudreaux, Nikolaev, & Holcombe, 2018; Sobel, 2008). As Acemoglu & Johnson (2005) explain, pro-market institutions involve lower risks of government expropriation and better enforcement of contracts. When institutions are weak, the risk of arbitrary expropriation reduces the benefits of entrepreneurial action, encourages opportunistic behavior, and reduces the expected return from new ventures (Estrin et al., 2013). Pro-market institutions help to minimize the risks associated with uncertainty, enabling entrepreneurial action (McMullen & Shepherd, 2006).

It is thus well-established that national economic freedom facilitates voluntary exchange, thereby increasing economic dynamism and development through a variety of outcomes (Hall & Lawson, 2014). This has the effect of increasing demand, facilitating firm growth aspirations (Estrin et al., 2013), and incentivizing in-migration (Ashby, 2010). Taken together, these factors naturally suggest a positive relationship between economic freedom and job creation.

Yet, *regional* economic freedom has received less attention in relation to such benefits (Hall, 2013). While the composition fallacy suggests caution in extrapolating to different levels of analysis (Caballero, 1991), there are strong reasons to expect regional economic freedom to also be positively related to net job creation. The relationship between economic freedom and in-migration appears to hold at the regional level (Ashby, 2007; Mulholland & Hernández-Julián, 2013); so do the benefits of regional economic freedom for development (Ashby, Bueno, & Martinez, 2013), entrepreneurship (Kreft & Sobel, 2005), well-being (Belasen & Hafer, 2013), and wages (Ashby et al., 2013). Economic freedom is thus a crucial feature of the regional institutional environment and an important antecedent of job creation, leading us to hypothesize the following:

**Hypothesis 2:** Regional economic freedom is associated with more job creation (or less job destruction).

### 2.3. Market preserving federalism

So far, our discussion has been broadly consistent with the standard framework of New Institutional Economics: macro-level institutions shape the incentives and opportunities of micro-level actors (Boudreaux & Nikolaev, 2018). While valuable, this framework runs the risk of oversimplification, because it tends only to incorporate two levels at a time (cf. Williamson, 2000). Recently, institutional accounts have been enriched by going beyond a two-level, macro-to-micro framework. Once the relevance of multiple institutional levels, rather than a single plane of institutions, is considered, the interactions among institutions become notable. For instance, meso-level structures may affect the causal linkages among political rules and individual actors (Kim et al., 2016). This logic has led to the insight that informal institutions like social trust or networks may offset the effects of legal institutions (Estrin et al., 2013; Kim & Li, 2014). In general, when multiple institutional levels have been treated concurrently, formal institutions have been conceptualized as a single level (Williamson, 2000). We enrich this story by looking *within* the formal institutional "level"—which is in fact comprised of multiple, hierarchical levels of government entities.

While research indicates a connection between regulation and reduced economic activity (Klapper et al., 2006), it is likely that this relationship is context dependent. In other words, we expect that other features of the institutional environment moderate the strength of this relationship. In theorizing about the interaction of formal governance institutions, we draw on insights from Constitutional Political Economy[3] and the theory of *market-preserving federalism* (Weingast, 1995) to identify the costs and benefits associated with a variety of forms of governance.

The theory of market-preserving federalism seeks to address a core political dilemma: how can government simultaneously be strong enough to protect individual rights and also be trusted not to use that strength to itself infringe on those rights? The question speaks to the tradeoff between the capacity to protect and the potential to oppress. If government is capable of preserving the wealth of its citizens, it is likely also capable of confiscating that wealth. Since the risk of confiscation can deter entrepreneurial activity (North, 1990), economic performance depends on the emergence of institutional arrangements where government protects and enforces property rights—facilitating the creation of wealth by individuals—but is also limited in the tendency to appropriate the wealth created by those individuals (Acemoglu & Johnson, 2005).

The federalist governance structure has long been heralded as a relatively successful mechanism for navigating this tightrope of political power (Hayek, 1960). Federalism is a form of decentralized governance characterized by hierarchical, autonomous governing entities, each with a clearly delineated scope of authority (Riker, 1964). According to Weingast (1995), *market-preserving* federalism exists when the regional government in a federal governance structure features three characteristics. First, regional governments, rather than national, must be the principal source of economic activity governance. Second, regional governments must not be able to substantively restrict trade with other regional units; there must be a "common market" at the national level. Third, the regional government must face a "hard budget constraint," meaning that it cannot borrow indefinitely or print money (Weingast, 1995, p. 4). When these criteria are met, the state is said to "credibly commit" to the preservation of market incentives: entrepreneurs can have a reasonable expectation that the government will enable their market activity without extensive confiscation (Qian & Weingast, 1997).

---

[3] Constitutional Political Economy focuses on the functioning and implications of alternative political institutional arrangements (Brennan & Buchanan, 2008). The term "constitutional" refers to a focus on the design and selection of constraints (Buchanan, 1990). For more information on this topic, we refer the reader to the seminal work of Buchanan & Tullock (1962) and to Persson & Tabellini (2005) for empirical examples.

The theory of market-preserving federalism emphasizes the benefits of this structure for economic development in comparison to other national economic systems. Indeed, nations that embrace the features of market-preserving federalism tend to enjoy economic development relative to those that do not (Weingast, 1995).[4]

## 2.4. *The moderating role of economic freedom*

The theory of market-preserving federalism informs our hypothesis about the interplay of national and regional government for regional economic activity. The logic suggests that a core benefit of regional enterprise policy is the ability to counteract the costs of national policy in the regional economy. Regional government thus plays the critical role of limiting the imposition of rules by the national government when those rules are inconsistent with regional economic needs. This follows from the idea that regional government is "closer" to the local citizenry and thereby has better access to knowledge about the regional economic system (Hayek, 1960). It is unlikely that a single set of institutional rules will be "optimal" for every regional economy (Dilli et al., 2018)—especially in light of the economic theory of regulation outlined above. This suggests that federal regulation need not accord with economic development in the heterogeneous regions affected by it. By contrast, regional policymakers are taking increasing responsibility for economic performance, as evidenced by widespread interest in the creation of industrial districts (Digiovanna, 1996; Tomlinson & Branston, 2017). With this in mind, regional policymakers may seek to offset the influence of federal regulation through their own policy decisions.

---

[4] It is important to note that the *de facto* governance structure can be federalist and market-preserving regardless of the *de jure*, formally designated institutions (Williamson, 1994). For instance, the late 20th century trend of marked economic development in China is said to have been facilitated by the government's adoption of a functional structure according with the features of market-preserving federalism (Qian & Weingast, 1996). Similarly, the Industrial Revolution-era United Kingdom and the 19th and 20th century United States experienced significant economic progress under market-preserving federalism (Weingast, 1995).

Economic freedom is a vehicle by which regional enterprise policy can be crafted to facilitate entrepreneurship and innovation. In addition to the direct benefits already explicated, an important, indirect benefit of regional economic freedom is the reduction of federal regulatory costs in the region. Whereas regulation imposes restrictions and thereby compliance costs, economic freedom facilitates experimentation and exchange. This is one way that regional institutions may be seemingly "incoherent" with national institutions but may yet be complementary (Schröder & Voelzkow, 2016). Case studies have illustrated how this kind of regional divergence from national regulation can facilitate economic activity for sectoral clusters with specific resource needs—e.g., a flexible and high-skilled workforce (Crouch & Voelzkow, 2009).

The moderating effect of regional economic freedom on national regulation for job creation emerges from this logic. Whereas national regulation imposes compliance costs and thus raises the costs of job creation, high levels of economic freedom can attenuate these costs. To the extent that a comprehensive governance structure consistent with economic freedom facilitates exchange, consumer demand will be increased. Thus, the relative benefits of bringing on additional labor— even that which requires additional regulatory training and compliance—are greater. Furthermore, regional institutions commonly conform to national institutional standards (Crouch, Gales, Trigilia, & Voelzkow, 2001). This means that national and regional policy are often complementary in their effects, so higher levels of economic freedom can represent a *departure* from national institutional standards. High levels of economic freedom enable businesses to grow and reduce regulatory barriers.

On the other hand, where regional economic freedom is low, the federalist governance structure is less likely to attenuate the impact of national regulation. With a more limited set of opportunities resulting from low economic freedom, entrepreneurs will remain encumbered by national regulation, relatively un-tailored to the regional economic climate. Furthermore, reductions of economic freedom typically entail regulatory restrictions at the regional level, which may complement national regulatory rules (Macey, 1990). The overlapping and cumulative nature of policies across government levels may be significant (Revesz, 2001). Low levels of regional economic freedom often involve high business taxation, adding another cost to employment growth alongside regulatory constraints.

The above discussion leads us to expect that regional economic freedom serves as a market-preserving mechanism relative to federal regulation: regions that are more economically free preserve regional economic activity by promoting labor market flexibility and insulating local enterprise from federal rulemaking. Altogether, this logic suggests that the costs of federal regulation for job creation are reduced where regional economic freedom is higher. This leads us to predict the following:

**Hypothesis 3:** Economic freedom will moderate the relationship between regulation and job creation such that regulation is less harmful to job creation as economic freedom increases.

We also note that the various components of economic freedom often have heterogeneous effects (Bjørnskov & Foss, 2008, 2016; Estrin et al., 2013; Heckelman & Stroup, 2005). For instance, the government spending component of economic freedom is weakly negatively correlated with the rest of the index at the country level (Heckelman & Stroup, 2005). Moreover, an examination of the area components of economic freedom finds that a summary index can mask nuanced relationships if the components affect economic activity in different directions (Carlsson & Lundström, 2002). For these reasons, it behooves us to consider how each component of the economic freedom index (government spending, taxation, and labor market freedom) moderates the relationship between regulation and net job creation. We describe the components of economic freedom in the data section and examine each component in our results section.

### 2.5. Moderation and the firm age spectrum

Integrating the theory of market-preserving federalism with the economic theory of regulation is a natural fit, as both take an economic and institutional view of competing interests in the formation of policy. Together, the theories suggest that regional economic freedom offsets the harms of national regulation for job creation. However, recent studies have also emphasized firm age as an important factor for job creation (Haltiwanger et al., 2013). We thus extend our theory to account for differential effects across the firm age spectrum.

We expect the moderating relationship to be most pronounced for young firms. To see this, consider that the economic theory of regulation regularly pits incumbents against entrants. Stigler's seminal work suggests that the 'industry,' comprised of firms already organized into a concentrated interest, 'demands' regulation in order to deter competitors from entering (Stigler, 1971). The implication is that regulation is most costly to potential and/or young firms—a prediction supported in the literature (Bertrand & Kramarz, 2002; Branstetter et al., 2014).

Conversely, the institution of economic freedom is said to be a critical enabler of entrepreneurship (Bjørnskov & Foss, 2016; Boudreaux, 2014; Bradley & Klein, 2016; Kreft &

Sobel, 2005). Economic freedom allows for flexibility and experimentation, creating favorable conditions for new ventures to be birthed and to grow. Together, this suggests that the interaction among regulation and regional economic freedom should be most prominent among entrepreneurial firms. Because regulation is often coopted by special interests, it is said to be particularly harmful to entrepreneurs; thus, the benefits of the state-level economic freedom that would offset these harms should accrue largely to new ventures.[5]

Nonetheless, we do expect the moderating relationship to hold for mature firms as well. Economic freedom can enable innovation and growth for all firms—not just new entrants. As our market-preserving federalism theory suggests, regional policy can safeguard for regional economic interests relative to national or extra-regional interest groups. Thus, there may also be disparate interests among industry participants across regions; mature firms will enjoy market-preserving benefits as well. Incumbents, however, are established and thus likely better able to organize to combat interregional political competition than are young firms—likely through regional policies not captured by economic freedom. By contrast, regional economic freedom tends to 'level' the intraregional playing field with respect to national regulation, mitigating cronyism and the need for firm experience with policymakers and thereby reducing the obstacles to competitive advantage for entrants. So while the market-preserving benefits of regional economic freedom are broadly applicable, our theory points to heterogeneity in the magnitude of these benefits with respect to firm age. Thus, we also hypothesize:

**Hypothesis 4:** Economic freedom will moderate the relationship between regulation and job creation to a greater extent for young entrepreneurial firms relative to mature, established firms.

3. **Data and Methods**

---

[5] While we do not ascribe to the view that equates new or young firm activity with entrepreneurship, we acknowledge these as a manifestation of entrepreneurship.

To operationalize our research questions, we construct an industry-county panel in the U.S. from 2003 to 2015 using several sources. For national regulation, we use *RegData*, a novel measure of regulatory stringency that quantifies the number of restrictive words (e.g., "shall," "must") in the text of the U.S. Code of Federal Regulations and uses a machine-learning algorithm to assign restrictions to industries.

For regional policy, we utilize a comprehensive measure of the enterprise policy climate: state economic freedom, as measured by the Frasier Institute's *Economic Freedom of North America* (EFNA) index. Whereas specific policy initiatives to foster entrepreneurship often fail (Acs, Astebro, Audretsch, & Robinson, 2016; Lerner, 2009), economic freedom—including small government, favorable tax policy, and flexible labor market policy—has a robust, positive relationship to entrepreneurial activity and employment growth across nations (McMullen et al., 2008; Nikolaev, Boudreaux, & Palich, 2018; Nyström, 2008) and regions (Calcagno and Sobel 2014; Gohmann et al. 2008; Sobel 2008).

Because we are interested in the effect of regulation on high-growth entrepreneurship outcomes, we use the Census Bureau's Quarterly Workforce Indicators (QWI) to measure our outcome of net job creation (Birch, 1987; Davis, Haltiwanger, & Schuh, 1996; Haltiwanger, Jarmin, & Miranda, 2013). Job creation is a key feature of economic dynamism and is thus the object of much scholarly attention (Decker et al., 2014) and a focal point for regional enterprise policy. To that end, QWI is an ideal measure for our purpose, as it reports employment dynamics by firm age within each major industry at a local geographic level. Positive net job creation indicates the number of jobs created exceeds the number of jobs destroyed whereas negative net job creation indicates the reverse.

### 3.1. Dependent variable: Net job creation

We use employment data from the quality workforce indicators (QWI) to construct our job creation measure. QWI data is available at the county-level for each North American Industry Classification System (NAICS) industry, and we use the two-digit NAICS level of classification. QWI provides these data for each quarter and in several age bins including: 1) 0–1 years, 2) 2–3 years, 3) 4–5 years, 4) 6–10 years, and 5) 11 or more years. To construct our job creation measure, we aggregate quarterly data to annual data for each industry-county observation, allowing comparison with our other variables. QWI provides several employment measures including end of year employment, job gains, and job losses. We combine the last two measures to create a net job gain (or loss) measure, which we denote as *net job creation*. Positive numbers for our measure indicate job creation (job gains > job losses); negative numbers indicate job destruction (job gains < job losses).

### 3.2. Independent variables

#### 3.2.1.  National Regulation

Our measure of regulation is gathered from RegData (McLaughlin & Sherouse, 2016). This dataset quantifies the number of regulatory restrictions in the Code of Federal Regulations, the federal administrative code in the U.S., from 1970 to 2017. RegData features an annual measure of industry regulation available at several industry levels (2- through 6-digit NAICS); we utilize the 2-digit NAICS in order to match with QWI data. This variable first measures the number of restrictive words—"shall," "must," "may not," "prohibited," and "required"—in each subsection ('part') of the Code of Federal Regulations. The dataset's authors then employ machine learning to assign a probability that the restrictions in a subsection apply to a given industry. The product of the restrictive words times the probability is calculated for each industry and then summed across all subsections for a given year, yielding a measure of federal regulation (Al-Ubaydli & McLaughlin, 2017). The dataset also reports which agencies issue the regulatory restrictions. Additionally, the dataset includes the total number of words rather than the restrictions as an alternative industry regulation measure. We use this alternative measure of regulation to test the robustness of our results.

### 3.2.2. *Regional Economic Freedom*

We use the economic freedom[6] measure from the Frasier Institute's *Economic Freedom of North America* (EFNA) index (Stansel, Torra, & McMahon, 2017). Economic freedom is comprised of nine variables in three areas[7]: (1) government spending, (2) taxes, and (3) labor market freedom. The first area, government spending, is measured as the extent of government consumption, transfers and subsidies, and insurance and retirement payments. The second area, taxes, is measured by income and payroll tax revenue, the top marginal income tax rate, the property tax rate, and sales tax revenues. The third area uses minimum wage legislation, government employment, and union density to measure labor market freedom. We use the subnational, state-level index, measured on a scale from zero (low economic freedom) to 10 (high economic freedom).

---

[6] See Berggren (2003) and Hall and Lawson (2014) for excellent reviews of the literature on economic freedom.
[7] We use the subnational index, which is the preferred index for comparisons within a single country (Stansel, Torra, & McMahon, 2017). An alternative measure is the 'all government' index, which is comprised of additional variables and areas. Refer to the Economic Freedom of North America report for more detail. https://www.fraserinstitute.org/studies/economic-freedom-of-north-america-2017

*3.3. Controls*

We include several variables to control for relevant county-level differences. To control for the health of the local economy, we use county median household income, unemployment rate (%), and poverty rate (%). We expect that counties with a healthy local economy will have more job creation. We also include several demographic controls. We use population and population density to control for agglomeration economies (Duranton & Puga, 2004; Porter, 1996; Rosenthal & Strange, 2004) and to proxy urban context (Griffith, 1981; McDonald, 1989). Lastly, we also include the number of firms as a proxy for the competitive density of the local area (Voss & Voss, 2008). A highly competitive market will likely lower profits, which might result in less firm growth and job creation. All controls are taken from the U.S. Census Bureau.

Our sample consists of 463,474 total observations. The sample selection is comprised as follows: we begin by collecting all available net job creation data from the QWI. This sample represents 1,043,924 county-industry-year observations. We then match these data with the EFNA index and lose 104,359 observations due to missing data. Next we match this dataset with the RegData and lose 280,655 observations due to RegData's omission of certain industries. Lastly, we match our dataset with the controls from the U.S. Census, which reduces our sample by 195,436 observations for a total of 463,474 observations. However, the number of observations range from 313,552 to 447,556 within the different firm-age categories due to data availability.

------------------------------
Insert Table 1 about here
------------------------------

Table 1 summarizes these data. On average, there are over 48 jobs created annually in a county. The average level of economic freedom is 7, which ranges from a low of 5.25 to a high of 8.46. There is substantial variation in industry-level regulations. The average number of restrictions for an industry is 60,247, which ranges from a low of 4,558 to a high of 209,220. The average household has a median household income of $42,849. Lastly, the average county has 89,347 residents or 175 persons per square mile (density), a poverty rate of 15.69 percent, an unemployment rate of 6.5 percent, and 209 firms. Table 2 provides a correlation matrix for these variables. Most variables are not highly correlated with other variables, which reduces concerns of multicollinearity. We now proceed to a discussion of our estimation methods.

------------------------------
Insert Table 2 about here
------------------------------

### 3.4. Estimation methods

To test our hypotheses, we model job creation as a function of several explanatory variables:

$$JC_{ijt} = \alpha + \beta_1 R_{ijt} + \beta_2 EF_{ijt} + \beta_3 (R_{ijt} \times EF_{ijt}) + X'_{ijt}\delta_{ijt} + \lambda_i + \theta_j + \Pi_t + u_{ijt} \quad (1)$$

The outcome variable, *JC*, denotes net job creation in county *i* industry *j* and year *t*. The right-hand side of the equation includes regulation (*R*), economic freedom (*EF*), their interaction, and a vector of controls (*X*) for each observation of county *i*, industry *j* in year *t*. The parameters, measured by *β*, capture the effect of each variable on job creation. In particular, $\beta_1$ and $\beta_2$ capture the direct effect of regulation and economic freedom whereas $\beta_3$ captures the effect of their interaction. The parameter *δ* captures the effects of each variable in the vector of controls (*X*). The parameters *λ*, θ, and *Π* capture county, industry, and year heterogeneity. These fixed-effects are included to control for common macroeconomic trends and unobserved regional and industry idiosyncrasies (Bournakis, Papanastassiou, & Pitelis, 2018). The parameter, *u*, is the disturbance term, which is assumed independently and identically distributed (iid). However, we control for potential heteroscedasticity by using robust standard errors clustered at the county-level. Except for state economic freedom (index), we express all explanatory variables in logs. This transforms all coefficient estimates into elasticities, easing the interpretation of our findings.

## 4. Results

### 4.1. Main results: The moderating effect is supported

Table 3 illustrates the direct effects of national regulation and regional economic freedom on net job creation as well as the moderating effect of regional economic freedom on the regulation-job creation relationship. Here, we test and find support for Hypotheses 1, 2 and 3. Model 1 presents our baseline results, which includes our control variables but does not include measures of regulation or economic freedom. Model 2 augments this model by including the measures of regulation and economic freedom. The results indicate that more regulation is associated with fewer net jobs created ($\beta = -14.12$; $p < 0.001$), while economic freedom is associated with more jobs created ($\beta = 31.12$; $p < 0.001$). Model 3 adds an interaction term to test our hypothesis that economic freedom moderates the effect of regulation. The likelihood ratio (LR) test that compares Models 2 and 3 indicates that the differences between the models are statistically significant, which suggests that the interaction model is appropriate. The results continue to support Hypotheses 1 and 2, while also providing support for Hypothesis 3. Specifically, Model 3 indicates that regulation is associated with fewer net jobs created ($\beta = -57.85$; $p < 0.001$), but this effect decreases as economic freedom increases ($\beta = 6.24$; $p < 0.001$). For instance, in states with the average level of economic freedom, a one percent increase in regulation is associated with 14 jobs destroyed.[8] However, for a one standard deviation increase in economic freedom, the negative effect of regulation on net job creation is attenuated by four jobs gained (four fewer jobs destroyed on net).[9]

-------------------------------
Insert Table 3 about here
-------------------------------

To better understand the moderating effect of regional economic freedom, we plot the marginal effects of regulation on net job creation at various levels of economic freedom along with 95 percent confidence intervals in Figure 1. The vertical axis denotes the effect on the number of jobs created on net whereas the horizontal axis denotes the quality of economic freedom; note that we restrict the action and prediction to the range of our sample rather than the range of the index (zero to ten) to avoid extrapolation. The figure illustrates that regulation has a more adverse effect on net job creation when economic freedom is lower. At the bottom of the economic freedom distribution ($EF = 5$), a one percent increase in regulation is associated with 20 fewer net jobs created *ceteris paribus*. However, in regions where economic freedom exceeds the average level ($EF \geq 7$), a one percent increase in regulation is associated with virtually no decrease in net jobs created. In fact, the effect of regulation is not statistically and significantly different from zero based on the standard 95 percent confidence interval at above-average levels of economic freedom.

-------------------------------
Insert Figure 1 about here
-------------------------------

### 4.2. Decomposing the EFNA index

---

[8] We arrive at this number by taking the average level of economic freedom (7) and multiplying it by the moderating effect (6.236), which is 43.65. We then subtract this number from the estimate for the log of restrictions (-57.85) and arrive at the difference of -14.2.

[9] The standard deviation of economic freedom is 0.62 and the moderating effect of economic freedom is 6.236. Therefore, a one standard deviation increase yields 3.87 (0.62 x 6.236) fewer jobs destroyed on net. We round this to four jobs.

Though we did not hypothesize about them individually, we also report results for each of the three EFNA sub-indices to investigate potential heterogeneity in the effects of different aspects of economic freedom. Models 4–6 of Table 3 present the decomposed results. Following standard practice (e.g., Wennberg, Pathak, & Autio, 2013), we replace the EFNA summary index with each of its three components (i.e., government spending, taxes, and labor market freedom) and include the interaction with each sub-index separately First, we note that regulation continues to be negative and statistically significantly harmful to net job creation, further supporting Hypothesis 1. Consistent with our results for the summary measure of economic freedom, we observe that tax freedom ($\beta = 28.79$; $p < 0.001$) and labor market freedom ($\beta = 32.17$; $p < 0.001$) moderate the effect of regulation on net job creation. Furthermore, we find positive and significant coefficients for the direct effects of these two economic freedom components. In contrast, we find no evidence to suggest that the freedom from government spending component moderates the effect of regulation on net job creation. This further validates the findings in previous studies that identify heterogeneity in economic freedom (Bjørnskov & Foss, 2016). Interestingly, we find that the direct effect of freedom from government spending on net job creation is negative and significant in each of Models 4–6.

### 4.3. Comparison of results by firm age

We now examine the effect of regulation on net job creation for different age firms in order to test Hypothesis 4. Table 4 separates our data into different firm age categories, which include young firms (less than one year of age) and mature firms (greater than 10 years of age). We focus on these two categories as they are the most consistent with our theoretical constructs and consistent with prior literature using QWI data (Curtis & Decker, 2018)[10]. Models 1–4 of Table 4

---

[10] We also estimated these models for the other age categories provided by the QWI dataset (2-3 years, 4-5 years, and 6-10 years). We found no statistical relationship for the age categories 2-3, 4-5, and 6-10 years of age. Moreover, we

detail the results for young firms (0-1 years of age); Models 5–8 detail the results for mature firms (≥ 11 years of age). In the main models, we find that the economic freedom summary index is associated with more net jobs created for young firms (Model 1; β = 4.12; p < 0.001) and mature firms (Model 5; β = 25.64; p < 0.001), but regulation is only associated with fewer net jobs created for mature firms (Model 5; β = –48.68; p < 0.001). Regulation is negative but insignificantly related to net job creation for young firms in our main model (Model 1; β = –9.18, p > 0.05). We do find negative and significant effects for regulation in each of the sub-index regressions for both young firms (Models 2–4) and mature firms (Models 6–8). Interestingly, economic freedom only moderates the effect of regulation on net job creation for mature firms: in Model 5, the coefficient on the main index interaction term is statistically significant and positive (β = 5.63; p < 0.001). We find continued support for the moderating effect for the taxation (Model 7; β = 5.88; p < 0.001) and labor market freedom (Model 8; β = 6.64; p < 0.001) sub-indices; as in the main results, the government size sub-index interaction is insignificant (Model 5; β = 1.50; p > 0.05). The young firm moderating terms are all insignificant: the interaction is economically and statistically insignificant in Model 1 (β = –0.45, p > 0.05), and the moderating relationship remains insignificant for each of the sub-indices (Models 2–4). Moreover, a t-test comparison between the estimates of young firms and old firms (models 1 vs 5; models 2 vs 6; models 3 vs 7; models 4 vs 8) indicates the estimates are statistically and significantly different (p < 0.001). Thus, we do not find support for Hypotheses 4.

-------------------------------
Insert Table 4 about here
-------------------------------

---

cannot think of a theoretically-sound reason to explain why regulation might affect each of these age categories differently. Therefore, we only examine and contrast the young age category (0–1 years) and mature age category (≥ 11 years) to be more consistent with theory and extant literature (Bailey & Thomas, 2017).

## 5. Robustness check

Table 5 tests the robustness of our main results by using an alternative measure of regulation. RegData provides this alternative measure, which reports the total *word count* for the regulations contained in the Code of Federal Regulations. The logic behind this alternative measure is that a lengthier administrative code naturally imposes more restrictions and limitations on firm activity. Consequently, larger word counts are indicative of more regulation. Model 1 presents baseline results with only control variables. Model 2 augments this model to include economic freedom and the alternate measure of regulation. The results indicate that economic freedom is associated with more net jobs created ($\beta = 31.11$; $p < 0.001$), but regulation has no direct effect on net job creation in the non-interaction model. Model 3 adds an interaction term to test our hypothesis that economic freedom moderates the effect of regulation on net job creation. The results support our core Hypotheses: regulation is associated with lower net job creation ($\beta = -43.51$; $p < 0.001$), but this effect decreases as economic freedom increases ($\beta = 5.52$; $p < 0.001$). We decompose the economic freedom index into its three areas in Models 4 through 6 of Table 5. The results hold for the tax freedom and labor market freedom components but not for the freedom from government spending component. This is precisely the result we found in Table 3, which further validates our findings.

------------------------------
Insert Table 5 about here
------------------------------

## 6. Discussion

We were motivated by the prevailing view in the entrepreneurship literature that presents regulation as a homogenous barrier to entrepreneurial job creation (Bradford, 2004; Djankov et al., 2002; Escribá-Pérez & Murgui-García, 2017; Klapper et al., 2006). Despite the large support, this view fails to account for the heterogeneous effects of regulatory institutions (Kim et al., 2016). Scholars have paid limited attention to the fact that regulatory costs are unequally dispersed, instead treating them as symmetrical across firms and regions. This approach is inconsistent with the stylized facts about entrepreneurship—some new ventures do emerge and grow even in highly regulated industries (Henrekson & Johansson, 2010). We argued that prior single-level and even multi-level studies cannot account for this because they overlook the multiple levels of government within which entrepreneurs are embedded. In many instances, entrepreneurs must navigate more than one formal institutional level and one informal institutional level; they must also operate within a formal institutional environment consisting of multiple political *layers*.

Our work suggests that modeling these multiple political layers is important because the impacts on entrepreneurial outcomes are contingent on one another. Thus, to examine why the effects of national regulation on entrepreneurship vary across regions, we proposed a model of *hierarchical institutional interdependence*—the effects of national institutional regulation are contingent upon the regional institutional framework. Drawing on the theory of market-preserving federalism (Weingast, 1995), which suggests that regional policymakers possess more local knowledge about their particular economic context, we predicted that regional institutions would oppose and counteract federal rules that do not accord with regional enterprise needs, yielding heterogeneity in the effects of federal regulation across regions.

Our analysis revealed that national regulation is negatively associated with entrepreneurial net job creation and that state economic freedom is positively associated with entrepreneurial net job creation. Critically, we uncovered a moderating effect where state economic freedom attenuates the adverse effect of regulation on entrepreneurial net job creation. The magnitude of this moderation is not trivial: a one percent increase in industry-level regulation was associated with 14 fewer jobs created on net, but this effect varied considerably with the quality of the regional institutional environment. In states that have very little economic freedom, a one percent increase in industry-level regulation was associated with roughly 27 fewer net jobs created in a county. On the other hand, regulation had no discernable effect on net job creation in states that have very high levels of economic freedom. This finding is notable—the economic costs of decreased net job creation brought about by national regulation are completely mitigated in the most economically free regions. Our analysis also revealed that these offsetting results are driven by older firms and not by young firms.

Our work speaks to the ongoing debate on the merits of a multilevel institutional framework such as the U.S. federalist system. Indeed, the "centralization versus decentralization" debate has a rich history, including the well-known arguments of Alexander Hamilton, James Madison, and John Jay in *The Federalist Papers*. More recently, a concern raised in political science is that voters governed by multilevel or overlapping political entities may have a more difficult time attributing economic outcomes to particular government levels, thereby mitigating electoral accountability (Anderson, 2006). While voters' ability to attribute responsibility for regional activity is beyond our scope, we nonetheless find that there are distinct and economically meaningful policy impacts for different institutional levels. Furthermore, we show that the regional institutional environment moderates the impact of national policy. To the extent that regional policy is more flexible and responsive to local economic conditions than national policy, the benefits to building the capacity of regional policymaking are greater than previously acknowledged.

While the relationships we observe are robust, our analysis also reveals nuance in the role of regional economic freedom—specifically in relation to the three sub-indices. The positive direct effect of regional economic freedom holds for the tax freedom and labor market freedom components. It does not hold, however, for the freedom from government spending component. In fact, the freedom government spending component has a negative direct relationship to net job creation. We also found that taxes and labor market freedom both moderated the effects of national regulation, but freedom from government spending did not. These results may be indicative of government spending's ability to foster short-run employment growth: greater government spending reduces a state's regional economic freedom score but may subsidize job creation. Of course, this does not imply that such subsidization is conducive to economic growth. Greater government spending will eventually impose costs, but they may be hidden or deferred through debt financing. We are unable to observe the long-term relationship between regional economic freedom (or its components) and net job creation, and while this is beyond the scope of our analysis, we believe it is worthy of future research. Our findings speak to the complexities of what makes a "good" institutional environment, and they affirm the value of addressing heterogeneity in the implications of various aspects of regional economic freedom (Dilli et al., 2018).

### 6.1. Limitations and future research suggestions

One critique of the institutional approach to regional economics is that regional institutions may be difficult to measure and hence to operationalize in analysis (Rodríguez-Pose, 2013). While the improvement and creation of regional institutional measures is indeed an object worthy of additional study (Teague, 2016), our work suggests that extant indices can yield fruitful insights about regional institutional variation. Furthermore, our findings are consistent with recent work showing that regional institutions can incentivize labor mobility, particularly among high-skilled workers (Mulholland & Hernández-Julián, 2013; Nifo & Vecchione, 2014). Extending this linkage, future research could consider to what extent the relationships we observe vary along the skill spectrum and how they relate to interregional migration. For instance, if federal regulations are less economically harmful to businesses in more economically free regions, does this further the incentive for firms to reallocate their activities to those regions?

Perhaps our most surprising result is that the relationships we observe appear to be driven by the oldest firms. The economic theory of regulation conceives of regulation as asymmetrically burdening potential entrants, thereby favoring incumbents (Stigler, 1971). Indeed, our *a priori* expectation was that this would translate to further reduce net job creation by young firms specifically, and that economic freedom would empower young firms asymmetrically. But the results we found are interesting in light of recent work addressing regulation and job creation across firm *size*. While some researchers have found negative effects of regulation for small firms (Bailey & Thomas, 2017), others find an equivocal relationship (Goldschlag & Tabarrok, 2018). We depart from these studies by considering firm age rather than size, which is arguably a better categorization for job creation (Haltiwanger et al., 2013). Future work might consider the effects of regulation while accounting for age and size concurrently; unfortunately, data constraints preclude this in the current exercise.

It is plausible that young firms' job turnover decisions are not primarily driven by regulation (Goldschlag & Tabarrok, 2018). It is also worth noting that net job creation for the median young firm is minimal (Decker et al., 2014). It is instead a small proportion of young firms that drive job creation; these "high-growth" young firms are not limited to a particular industry (Henrekson & Johansson, 2010). We do not directly test the impact of regulation on the incidence of high-growth entrepreneurship, instead looking at all young firms. Thus, for the typical observation in our young firm subsample, net job creation may simply be economically insignificant.

In contrast, state economic freedom does offset the negative effects of regulation among older firms—an important finding in light of the literature on regional employment growth (Acs & Armington, 2004; Digiovanna, 1996; Holm & Østergaard, 2015). Much of this work emphasizes the direct effects of regional policy in driving variation in net job creation. The evidence suggests that a variety of policy regimes may be conducive to net job creation in different contexts—e.g., those tailored either toward new ventures or incumbents (Audretsch & Fritsch, 2002). Interestingly, our findings position economic freedom as a vehicle for net job creation by incumbents.

While our analysis makes important headway into consideration of the interaction of governance institutions for employment and entrepreneurship, it does have limitations worth noting. First, it is worth considering the possibility of reverse causality among net job creation and our key independent variables, regulation and economic freedom. This logic argues that new regulations such as employee health and safety or technology standards might arise in response to job creation. Upon further consideration, however, this seems less plausible: because regulations are more closely tied to the political process and rent seeking interests (Murphy et al., 1993), it is far from obvious as to why regulation would substantively change in response to net job gains.

Second, measuring institutions is a complex endeavor (Rodríguez-Pose, 2013). Although we utilize the best available measures of federal and state level policies we consider, measurement error is a plausible issue. That said, similar studies using RegData do not find evidence of measurement error (Goldschlag & Tabarrok, 2018), which helps to alleviate this concern.

Third, our research focuses on the net outcomes of firm growth activity at the industry level. We believe this is beneficial in order to observe the economic implications of the entrepreneurial competition we have described, and we retain as local a unit of analysis as available data allow (i.e., the industry-county-year). However, research engaging the firm and individual levels that follows our hierarchical institutional interdependence framework has a great deal to offer. It is clear that the effects of regulation are also heterogeneous with respect to both firm characteristics (e.g., their resource base and capabilities) and entrepreneurs' cognitive traits (Boudreaux, Nikolaev, & Klein, 2018; Estrin et al., 2013). Explicit modeling of the multilevel formal institutional environment represents a promising direction for scholars developing the micro-foundations of the institutions-entrepreneurship nexus.

## 7. Conclusion

In light of the global shift toward the entrepreneurial economy (Thurik, Stam, & Audretsch, 2013), the accumulation of federal regulation becomes increasingly important. We explore how policies at different governance levels interact to influence regional enterprise. While regional policymakers may be unable to change national policy outcomes, their decisions can shape the impact of those policies on entrepreneurial job creation at the subnational level.

Using insights from the theory of market preserving federalism (Weingast, 1995), we have modeled the institutional hierarchy to capture how entrepreneurs are embedded within multiple institutional levels of governance. When lower levels of governance deviate from national levels, local policy makers have the ability to counteract decisions to improve the local economy through policies that enhance employment and net job creation. This insight has been overlooked in previous discourse.

Our work offers four implications. First, because scholars tend to focus on direct effects of regional institutions on regional economic activity, the importance of such institutions has likely been understated. The finding that regional economic freedom moderates the effects of national regulation suggests that researchers should consider both direct and indirect consequences of regional enterprise policy when formulating and testing models of the policy nexus. Second, our work implies that the impact of regulation on entrepreneurs is more nuanced than previously acknowledged. Not only have we identified regional heterogeneity in the effects of regulation, we also found surprising evidence that regulation's costs were not moderated for young firms, at least on the job creation margin. Third, our work points to regional economic freedom as an fruitful object of inquiry for the entrepreneurial ecosystems literature, as it represents an important answer to calls for policy reform that would foster an entrepreneurial regional economy (Stam, 2015). Finally, by incorporating the theory of market-preserving federalism, we demonstrate the rich potential and ready availability of explicitly multilevel theoretical frameworks from other disciplines that can be incorporated into regional studies.

**Table 1.**

Summary Statistics

| Variable | Obs. | Mean | Std. Dev. | Min | Max |
|---|---|---|---|---|---|
| Job creation | 463,474 | 48.77 | 478 | -27043 | 36651 |
| EFNA index | 463,474 | 6.99 | 0.62 | 5.25 | 8.46 |
| Restrictions | 463,474 | 60247 | 46040 | 4558 | 209220 |
| Median household income | 463,474 | 42849 | 11180 | 16868 | 125900 |
| Unemployment rate (%) | 463,474 | 0.065 | 0.027 | 0.01 | 0.24 |
| Population | 463,474 | 89347 | 243504 | 40 | 5330484 |
| Poverty (%) | 463,474 | 15.69 | 6.15 | 2.5 | 62 |
| Population density | 463,474 | 175 | 733 | 0.005 | 26179 |
| Number of firms | 463,474 | 209 | 1522 | 1 | 130454 |

**Table 2.**

Correlation Matrix

| | | [1] | [2] | [3] | [4] | [5] | [6] | [7] | [8] | [9] |
|---|---|---|---|---|---|---|---|---|---|---|
| Job creation | [1] | 1 | | | | | | | | |
| EFNA index | [2] | 0.007* | 1 | | | | | | | |
| Restrictions | [3] | -0.023* | 0.004* | 1 | | | | | | |
| Median household income | [4] | 0.101* | -0.01* | 0.025* | 1 | | | | | |
| Unemployment rate | [5] | -0.029* | -0.326* | 0.022* | -0.285* | 1 | | | | |
| Population | [6] | 0.339* | -0.036* | -0.007* | 0.280* | -0.002 | 1 | | | |
| Poverty percent | [7] | -0.032* | -0.002 | 0.027* | -0.681* | 0.492* | -0.093* | 1 | | |
| Population density | [8] | 0.199* | -0.06* | -0.001 | 0.217* | 0.014* | 0.576* | -0.06* | 1 | |
| Number of firms | [9] | 0.151* | -0.02* | -0.003 | 0.119* | -0.008* | 0.364* | -0.042* | 0.211* | 1 |

*Note.* * p<0.05.

**Table 3.**
Effects on job creation

| | (1) | (2) | (3) | (4) | (5) | (6) |
|---|---|---|---|---|---|---|
| | Dependent variable = job creation | | | | | |
| Median household income (log)[a] | -178.1*** | -178.7*** | -179.8*** | -175.9*** | -176.24*** | -176.40*** |
| | (21.75) | (21.74) | (21.71) | (21.87) | (21.89) | (21.95) |
| Unemployment (log) | -85.54*** | -77.95*** | -78.21*** | -78.42*** | -78.42*** | -78.76*** |
| | (8.117) | (7.706) | (7.715) | (7.69) | (7.69) | (7.70) |
| Population (log)[a] | 6.967 | 5.573 | 4.708 | 7.63 | 6.82 | 6.81 |
| | (11.94) | (11.70) | (11.62) | (11.32) | (11.23) | (11.26) |
| Poverty (log) | 36.87*** | 39.20*** | 39.44*** | 36.82*** | 37.28*** | 36.71*** |
| | (9.132) | (9.247) | (9.254) | (9.23) | (9.24) | (9.23) |
| Population density (log)[a] | 1.090 | 1.104 | 1.137 | 0.91 | 0.94 | 0.96 |
| | (3.530) | (3.506) | (3.503) | (3.50) | (3.50) | (3.50) |
| Number of establishments (log) | 0.548 | 0.549 | 0.549 | 0.53 | 0.53 | 0.531 |
| | (0.578) | (0.577) | (0.577) | (0.58) | (0.58) | (0.58) |
| Regulations | | | | | | |
| Log of restrictions (R) [a b] | | -14.12** | -57.85*** | -24.65** | -54.16*** | -72.91*** |
| | | (5.421) | (11.95) | (7.70) | (10.43) | (10.99) |
| Economic freedom | | | | | | |
| EFNA index | | 31.12*** | 30.79*** | | | |
| | | (9.342) | (9.322) | | | |
| EFNA government spending | | | | -12.00*** | -12.13*** | -12.22*** |
| | | | | (3.78) | (3.79) | (3.80) |
| EFNA taxes | | | | 33.82*** | 33.60*** | 33.87*** |
| | | | | (8.01) | (8.01) | (8.01) |
| EFNA labor market freedom | | | | 33.07*** | 33.11*** | 33.56*** |
| | | | | (8.24) | (8.24) | (8.28) |
| Moderating effects | | | | | | |
| EFNA index x R | | | 6.236*** | | | |
| | | | (1.856) | | | |
| EFNA government spending x R | | | | 1.37 | | |
| | | | | (1.07) | | |
| EFNA taxes x R | | | | | 6.27*** | |
| | | | | | (1.71) | |
| EFNA labor market freedom x R | | | | | | 8.36*** |
| | | | | | | (1.64) |
| Model fit statistics | | | | | | |
| Number of observations | 463474 | 463474 | 463474 | 463474 | 463474 | 463474 |
| Number of groups (counties) | 2698 | 2698 | 2698 | 2698 | 2698 | 2698 |
| AIC | 6950678 | 6950647 | 6950624 | 6950573 | 6950543 | 6950534 |
| Degrees of freedom | 31 | 33 | 34 | 34 | 34 | 34 |
| Prob > F | *** | *** | *** | *** | *** | *** |
| Log-likelihood | -3475307 | -3475289 | -3475277 | -3475250 | -3475234 | -3475230 |
| County-level variance | 191.6 | 194.6 | 195.4 | 193.5 | 194.2 | 194.1 |
| Model residual variance | 438.1 | 438.1 | 438.1 | 438.3 | 438.0 | 438.0 |
| % of variance, rho | 16.1 | 16.5 | 16.6 | 16.3 | 16.4 | 16.4 |
| LR test of rho = 0 [c] | *** | *** | *** | *** | *** | *** |
| LR test of model fit [d] | --- | --- | * | * | * | * |

*Note.* Dependent variable is annual number of jobs created. Standard errors are robust-clustered at the county-level and reported in parentheses. [a] denotes 1000s. [b] mean-centered. [c] Test of significance of county-level variance. [d] Likelihood ratio test that compares model 2 to models 3-6. Year and industry fixed effects included in all models.
* p<0.05
** p<0.01
*** p<0.001

**Table 4.**
Effects on net job creation (results stratified by firm age)

| | Dependent variable = Net job creation | | | | | | | |
| --- | --- | --- | --- | --- | --- | --- | --- | --- |
| | 0–1 years | | | | ≥ 11 years | | | |
| | (1) | (2) | (3) | (4) | (5) | (6) | (7) | (8) |
| Median household income (log)a | 2.593 (4.23) | 3.264 (4.11) | 3.247 (4.10) | 3.242 (4.10) | -151.6*** (19.76) | -148.7*** (19.93) | -148.9*** (19.95) | -148.9*** (20.00) |
| Unemployment (log) | -11.89*** (1.42) | -12.11*** (1.46) | -12.11*** (1.46) | -12.10*** (1.46) | -51.87*** (5.63) | -52.04*** (5.57) | -52.03*** (5.57) | -52.28*** (5.579) |
| Population (log)a | 9.244** (3.16) | 9.110** (3.15) | 9.172** (3.15) | 9.121** (3.15) | -6.730 (9.50) | -4.595 (9.06) | -5.342 (9.05) | -5.206 (9.085) |
| Poverty (log) | -2.029 (1.63) | -2.135 (1.63) | -2.165 (1.63) | -2.126 (1.63) | 36.93*** (7.88) | 35.06*** (7.87) | 35.52*** (7.88) | 34.97*** (7.869) |
| Population density (log)a | -2.465 (1.42) | -2.466 (1.42) | -2.471 (1.42) | -2.468 (1.42) | 4.772 (3.06) | 4.597 (3.05) | 4.632 (3.04) | 4.643 (3.044) |
| Number of establishments (log) | 0.110 (0.13) | 0.109 (0.13) | 0.109 (0.13) | 0.109 (0.12) | 0.170 (0.62) | 0.153 (0.62) | 0.158 (0.62) | 0.152 (0.624) |
| Regulations | | | | | | | | |
| Log of Restrictions (R) a b | -9.175 (4.722) | -10.88** (3.410) | -9.291* (3.984) | -9.942* (4.024) | -48.68*** (9.295) | -20.69** (6.38) | -46.64*** (8.126) | -55.79*** (9.276) |
| Economic freedom | | | | | | | | |
| EFNA index | 4.123** (1.333) | | | | 25.64** (7.814) | | | |
| EFNA government spending | | 0.224 (0.729) | 0.215 (0.727) | 0.212 (0.726) | | -6.990* (3.056) | -7.080* (3.060) | -7.132* (3.068) |
| EFNA taxes | | 3.162** (1.102) | 3.215** (1.097) | 3.161** (1.103) | | 25.08*** (6.567) | 24.84*** (6.567) | 25.12*** (6.566) |
| EFNA labor market freedom | | 1.437 (1.479) | 1.436 (1.477) | 1.445 (1.472) | | 24.33*** (6.950) | 24.37*** (6.949) | 24.68*** (6.989) |
| Moderating effects | | | | | | | | |
| EFNA index x R | -0.449 (0.671) | | | | 5.632*** (1.277) | | | |
| EFNA gov't spending x R | | -0.182 (0.370) | | | | 1.503 (0.767) | | |
| EFNA taxes x R | | | -0.493 (0.640) | | | | 5.879*** (1.172) | |
| EFNA labor market freedom x R | | | | -0.342 (0.544) | | | | 6.637*** (1.222) |
| Model fit statistics | | | | | | | | |
| Number of observations | 341063 | 341063 | 341063 | 341063 | 447556 | 447556 | 447556 | 447556 |
| Number of groups (counties) | 2698 | 2698 | 2698 | 2698 | 2698 | 2698 | 2698 | 2698 |
| AIC | 4136467 | 4136468 | 4136466 | 4136468 | 6526453 | 65264267 | 6526389 | 6526392 |
| Degrees of freedom | 34 | 36 | 36 | 36 | 34 | 36 | 36 | 36 |
| Prob > F | *** | *** | *** | *** | *** | *** | *** | *** |
| Log-likelihood | -2068198 | -2068197 | -2068196 | -2068197 | -3263191 | -3263176 | -3263157 | -3263159 |
| County-level variance | 68.82 | 68.89 | 68.85 | 68.88 | 98.94 | 97.71 | 98.34 | 98.20 |
| Model residual variance | 104.5 | 104.5 | 104.5 | 104.5 | 356.1 | 356.1 | 356.1 | 356.1 |
| % of variance, rho | 0.303 | 0.303 | 0.303 | 0.303 | 0.0717 | 0.0700 | 0.0709 | 0.0707 |
| LR test of rho = 0 c | *** | *** | *** | *** | *** | *** | *** | *** |

*Note.* Dependent variable is annual number of jobs created. Standard errors are robust-clustered at the county-level and reported in parentheses. A t-test comparison between the estimates of young firms and old firms indicates the estimates are statistically and significantly different (p<0.001). This test compares models 1 vs 5, models 2 vs 6, models 3 vs 7, and models 4 vs 8. a denotes 1000s. b mean-centered. c Test of significance of county-level variance. Year, county, and industry fixed effects included in all models.
* p<0.05
** p<0.01
*** p<0.001

**Table 5.**
Effects on net job creation (alternative measure of regulation)

| | Dependent variable = Net job creation | | | | | |
|---|---|---|---|---|---|---|
| | (1) | (2) | (3) | (4) | (5) | (6) |
| Median household income (log) | -178.1*** | -178.8*** | -179.9*** | -175.84*** | -176.35*** | -176.72*** |
| | (21.75) | (21.74) | (21.72) | (21.86) | (21.91) | (21.97) |
| Unemployment (log) | -85.54*** | -77.95*** | -78.24*** | -78.40*** | -78.42*** | -78.87*** |
| | (8.117) | (7.706) | (7.720) | (7.69) | (7.69) | (7.71) |
| Population (log) | 6.967 | 5.579 | 4.652 | 7.69 | 6.78 | 6.59 |
| | (11.94) | (11.71) | (11.60) | (11.21) | (11.21) | (11.22) |
| Poverty (log) | 36.87*** | 39.20*** | 39.45*** | 36.81*** | 37.31*** | 36.69*** |
| | (9.132) | (9.247) | (9.249) | (9.23) | (9.24) | (9.24) |
| Population density (log) | 1.09 | 1.100 | 1.137 | 0.91 | 0.94 | 0.98 |
| | (3.53) | (3.506) | (3.503) | (3.50) | (3.50) | (3.50) |
| Number of establishments (log) | 0.548 | 0.549 | 0.548 | 0.53 | 0.53 | 0.531 |
| | (0.578) | (0.577) | (0.577) | (0.58) | (0.58) | (0.58) |
| Regulations | | | | | | |
| Log of words (W) | | -4.584 | -43.51*** | -10.38 | -39.29*** | -65.79** |
| | | (3.471) | (12.49) | (7.67) | (10.56) | (10.95) |
| Economic freedom | | | | | | |
| EFNA index | | 31.11*** | 30.75*** | | | |
| | | (9.342) | (9.313) | | | |
| EFNA government spending | | | | -11.98*** | -12.15*** | -12.30*** |
| | | | | (3.78) | (3.79) | (3.80) |
| EFNA taxes | | | | 33.82*** | 33.59*** | 33.85*** |
| | | | | (8.01) | (8.00) | (8.01) |
| EFNA labor market freedom | | | | 33.05*** | 33.13*** | 33.73*** |
| | | | | (8.23) | (8.24) | (8.28) |
| Moderating effects | | | | | | |
| EFNA x W | | | 5.519** | | | |
| | | | (1.931) | | | |
| EFNA government spending x W | | | | 0.75 | | |
| | | | | (1.11) | | |
| EFNA taxes x W | | | | | 5.47***** | |
| | | | | | (1.80) | |
| EFNA labor market freedom x W | | | | | | 8.61*** |
| | | | | | | (1.66) |
| Model fit statistics | | | | | | |
| Number of observations | 463474 | 463474 | 463474 | 463474 | 463474 | 463474 |
| Number of groups (counties) | 2698 | 2698 | 2698 | 2698 | 2698 | 2698 |
| AIC | 6950678 | 6950649 | 6950630 | 6950577 | 6950551 | 6950530 |
| Degrees of freedom | 31 | 33 | 34 | 34 | 34 | 34 |
| Prob > F | *** | *** | *** | *** | *** | *** |
| Log-likelihood | -3475307 | -3475290 | -3475280 | -3475251 | -3475239 | -3475228 |
| County-level variance | 191.6 | 194.7 | 195.5 | 193.45 | 194.2 | 194.3 |
| Model residual variance | 438.1 | 438.1 | 438.1 | 438.0 | 438.0 | 438.0 |
| % of variance, rho | 16.1 | 16.5 | 16.6 | 16.3 | 16.4 | 16.4 |
| LR test of rho=0 | *** | *** | *** | *** | *** | *** |
| LR test of model fit | --- | --- | * | * | * | * |

*Note.* Dependent variable is annual number of jobs created. Standard errors are robust-clustered at the county-level and reported in parentheses. [a] denotes 1000s. [b] mean-centered. [c] Test of significance of county-level variance. [d] Likelihood ratio test that compares model 2 to models 3–6. Year and industry fixed effects included in all models.

\* p<0.05
\*\* p<0.01
\*\*\* p<0.001

**Figure 1.**
Marginal Effect of Regulation on Net Job Creation with 95% Confidence Intervals

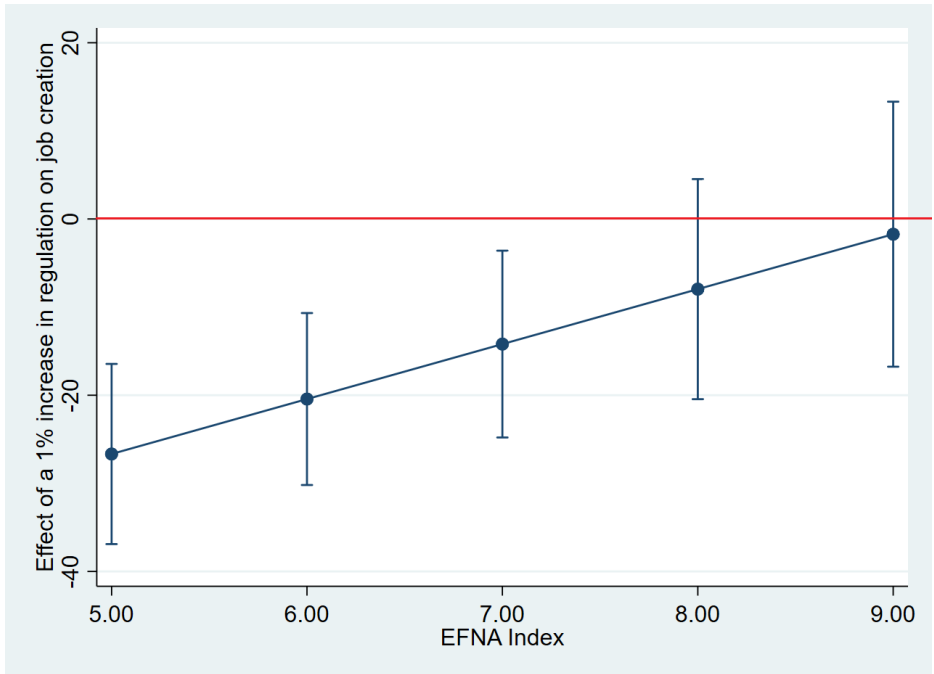